# A Telecom Perspective on the Internet of Drones: From LTE-Advanced to 5G

Guang Yang*, Xingqin Lin#,Yan Li*, Hang Cui*, Min Xu*, Dan Wu*, Henrik Rydén#, Sakib Bin Redhwan#

* China Mobile Research Institute, # Ericsson

Contact: yangguangyj@chinamobile.com; xingqin.lin@ericsson.com.

*Abstract—* **Drones are driving numerous and evolving use cases, and creating transformative socio-economic benefits. Drone operation needs wireless connectivity for communication between drones and ground control systems, among drones, and between drones and air traffic management systems. Mobile networks are well positioned to identify, track, and control the growing fleet of drones. The wide-area, quality, and secure connectivity provided by mobile networks can enhance the efficiency and effectiveness of drone operations beyond visual line-of-sight range. In this article, we elaborate how the drone ecosystem can benefit from mobile technologies, summarize key capabilities required by drone applications, and analyze the service requirements on mobile networks. We present field trial results collected in LTE-Advanced networks to gain insights into the capabilities of the current 4G+ networks for connected drones and share our vision on how 5G networks can further support diversified drone applications.**

## I. INTRODUCTION

Drone-related applications are flourishing [1]. Drone industry is moving into fast lane with accelerated involvements across industries [2]. Major trends in multi-directional drone applications include intelligent flight, broadband transmission, and diversified functions, leading towards the Internet of Drones [3][4]. Civilian drones mainly find applications in consumer entertainment uses and industrial uses such as inspection, agriculture, logistics, monitoring, and rescue. Entertainment uses of drones include live broadcast of large-scale events (e.g., sports events and music concerts), film-making, and ad-shooting. With the advantages of low cost, high flexibility, high security, less-affected by natural environment and better visual perspective, drones are more and more widely used in the field of infrastructure inspection. Agriculture is another important industry area where drones can help boost production (e.g., spraying of agricultural chemicals and mapping of farmland information). Drones for logistics and distribution can save time, reduce costs, and save manpower.

With the rapid growth of the drone industry and expansion of drone applications, limitations of existing solutions for point-to-point communication and control between a drone and its controller have become increasingly obvious. There are several major problems that the drone industry is currently faced with: 1) limited operation range, i.e., most drones can only fly within the visual line-of-sight (LOS) distances of the controllers; 2) limited bandwidth that cannot guarantee real-time high-definition (HD) image/video transmission; 3) inaccurate tracking, i.e., existing positioning based on global navigation satellite system (GNSS) may not be reliable due to potential spoofing and jamming; and 4) limited operation time due to battery constraints.

Mobile networks can greatly extend the drone application fields and help generate tremendous economic value [5][6]. Long-term evolution advanced (LTE-Advanced, a.k.a., 4G+) and 5G mobile networks help to overcome the aforementioned obstacles faced by the drone industry through providing capabilities such as remote and real-time control, HD image/video transmission, efficient drone identification and regulation, and high-precision positioning. Mobile networks can greatly extend the communication range for real-time transmission of remote drone control information compared to using proprietary communication technologies that are not deployed nationwide. High speed, low latency communications provided by mobile networks enable HD image/video transmission and thus lead to better user experience.

In this article, we elaborate how the drone ecosystem can benefit from mobile technologies in Section II. We then in Section III summarize key capabilities required by drone applications and analyze the service requirements on mobile networks including data rate, latency, regulation, and positioning. In Section IV, we present field trial results collected in LTE-Advanced networks to gain insights into the capabilities of the current 4G+ networks for connected drones. We find that the existing LTE-Advanced networks targeting terrestrial usage can support the initial deployment of low altitude drones, but there may be challenges in meeting high speed and low latency requirements associated with certain drone applications.

5G networks will introduce new technologies to provide 3D coverage enhancement, high data rate, customized end-to-end quality-of-service (QoS) guarantee, and efficient identification and monitoring based on big data analysis. In Section V, we share our vision on how 5G networks can further support diversified drone applications. In particular, we give a concrete example to illustrate how the collected data may be exploited by 5G networks using machine learning methods to assist with drone identification. In Section VI, we conclude by pointing out some fruitful avenues pertinent to mobile networks for future research.

## II. FOUR REASONS WHY DRONE ECOSYSTEM SHOULD TAKE ADVANTAGE OF MOBILE TECHNOLOGIES

Drone operation needs wireless connectivity for communication between drones and ground control systems, among drones, and between drones and air traffic management systems. Mobile networks are well positioned to identify, track, and control the growing fleet of drones [7]. The drone

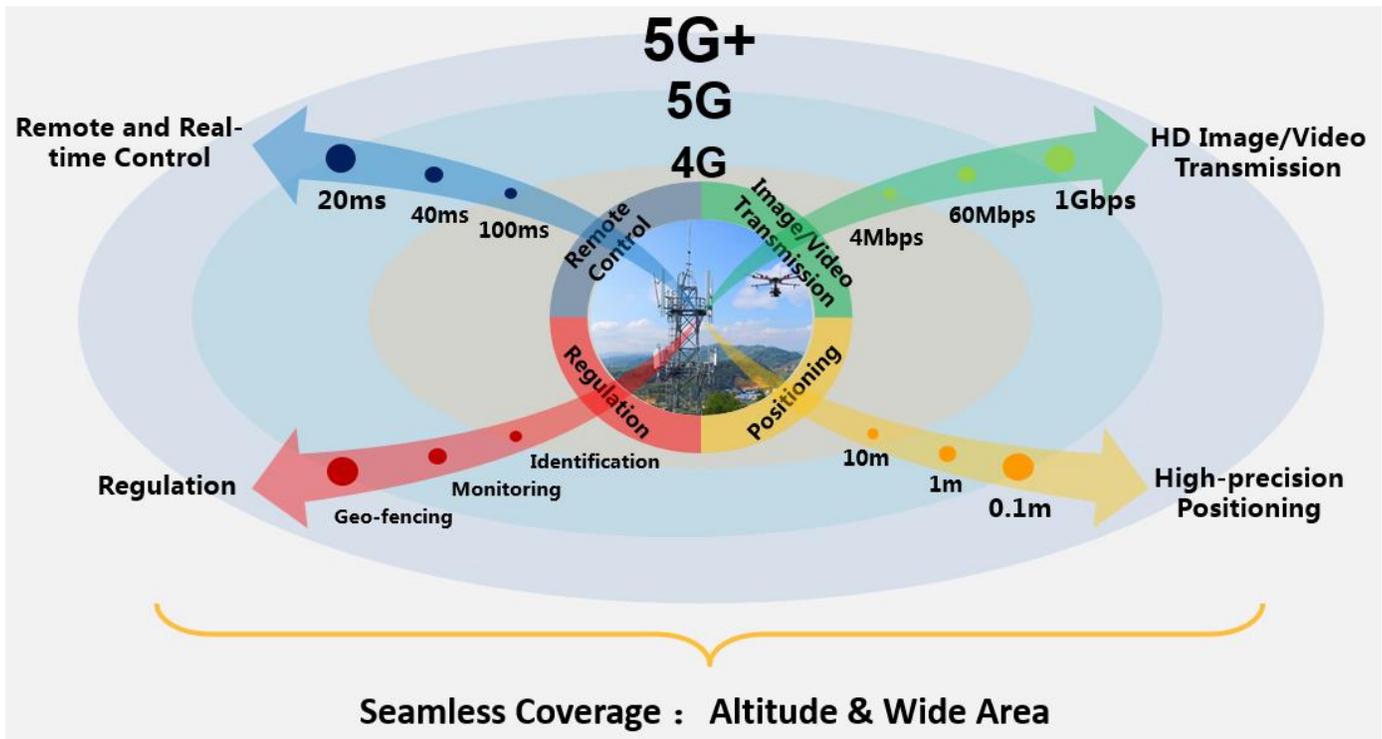

**Figure 1: Mobile networks stand ready to support drone uses and applications.**

ecosystem should take advantage of the mobile technologies for drone operations.

**1) Mobile networks stand ready nationwide to power flying drones beyond ground devices.** Nationwide network coverage is needed to safely expand low altitude drone operations into the national airspace. Developing a clean slate technology and rolling out a new nationwide network would require extensive investment in research, product development, testing, field trials, and infrastructure. These would lead to long time to market and may not be economically viable. Mobile networks are already up and running, and they provide nationwide coverage. The third generation partnership project (3GPP), which is a global collaboration between groups of telecommunications associations for developing and maintaining system specifications for mobile technologies, has dedicated a significant effort during its Release 15 to study LTE connected drones and concluded that it is feasible to use LTE networks to provide connectivity to low altitude drones [8][9].

**2) Mobile networks provide wide-area, quality, and secure connectivity for drone operations.** Many drone use cases require beyond visual LOS communications that can be made possible by the wide-area connectivity of mobile networks. The drone use cases are rapidly growing and expanding, covering a wide variety of industries and services, as described in Section I. With LTE-Advanced and 5G technologies, mobile networks can offer quality services featured with high mobile broadband data rate, low latency, large system capacity and robust reliability. These quality connectivity services are essential foundations for the fast growth of the drone applications that come with diverse requirements. The mobile connections are encrypted and secure, which can help drone operators meet the highest security standards.

**3) Mobile networks can assist with drone identification, authorization, and geo-fencing.** Mobile networks are equipped with a variety of tools to identify and authorize users and devices that can access the networks [10]. The International Mobile Equipment Identity (IMEI) is used to identify the mobile device, and the International Mobile Subscriber Identity (IMSI) stored on a Subscriber Identification Module (SIM) card is used to identify the user in a mobile network. These two secure technologies can be used for the operation of drones, with the IMEI being used for drone device identification and the IMSI on a SIM card being used for drone operator registration. The secure connectivity provided by mobile networks can also support other drone identification mechanisms such as certificate-based schemes.

No-fly zone is an important air traffic regulation that prohibits flights in a specific region of airspace. For the enforcement of no-fly zones, drones are required to be equipped with electronic geo-fencing functions. A drone cannot take off or will land once it detects that it is in a no-fly zone based on GNSS positioning. Unfortunately, the current electronic geo-fencing functions may be vulnerable due to the potential GNSS spoofing, jamming, and/or software cracking. Mobile networks can assist with geo-fencing to improve the robustness of electronic geo-fencing functions. Mobile networks based positioning can be used to locate the drones or counteract GNSS spoofing and jamming [11][12]. When a drone enters a no-fly zone, an alert can be sent via mobile networks to the drone operator, trusted enforcement authority, or drone traffic management system so that appropriate actions can be taken. Mobile networks can also be used to communicate the timely information update of no-fly zones, e.g., the setup of a temporary no-fly zone in a stadium in the case of a large event.

**4) Mobile technologies are based on standards and are evolving, facilitating interoperability and vibrant global growth for the evolution of drone ecosystem.** The mobile industry has developed LTE-Advanced standards and is working on 4G+ evolutions and 5G standards. The developed global technology standards are based on industry-wide consensus, and they are evolving in a backward compatible manner. They provide a global, interoperable, and scalable platform for the drone ecosystem to benefit from economies of scale to develop innovative services. With mobile networks being the backbone of communications systems for drone operations, the solutions for communication, security, identification, authorization, geo-fencing, tracking, monitoring, controlling, and law enforcement can take advantage of the well-established mobile technology standards and mobile ecosystem.

### III. COMMUNICATION REQUIREMENTS OF DRONE USES AND APPLICATIONS

To support the diverse drone-related applications, we summarize five essential capabilities in this section: 1) seamless coverage, 2) remote and real-time control, 3) HD image/video transmission, 4) drone identification and regulation, and 5) high-precision positioning, as illustrated in Figure 1.

#### A. Seamless Coverage

Table 1 summarizes four levels of coverage requirements in both altitude and wide area coverage. For aerial entertainment, hot-spot coverage (stadiums, tourist areas and commercial areas) is sufficient. For power line or base station (BS) inspection and logistics, wide-area coverage in urban, suburban and rural area are needed. Seamless coverage for drones will become more important for network planning in the future. Different from traditional network coverage mainly serving ground users, improved coverage in the sky is required to serve drone users that fly at different altitudes. For vegetation protections (e.g., spraying of agricultural chemicals), coverage up to 10 m altitude is sufficient. For power line or BS inspection, coverage up to 50 m – 100 m altitude is needed. For mapping of farmland, coverage up to 200 m – 300 m altitude is needed, while for upper-air pipeline inspection coverage up to 300 m – 3000 m altitude may be required. Serving such wide range of coverage scenarios at different altitudes is challenging for mobile networks.

#### B. Remote and Real-time Control

Remote and real-time communication capability enables a remote controller to issue command and control instructions in time based on the real-time flight status report from drones, such as spatial coordinates and equipment status. Remote and real-time control is mainly used for flight condition monitoring, drone task and equipment management, and emergency control. To enable remote control for drones, certain *data rate and latency* requirements should be met. In terms of data rate, the downlink (from BS to drone) data rate requirement is about 300-600 kbps in many application scenarios, and the current 4G+ networks can meet this requirement. For the future applications, such as remote real-time operation, the latency requirement will be stringent to guarantee the operation accuracy and experience.

Table 1 summarizes four levels of end-to-end latency and network latency requirements. For image/video transmission, the end-to-end latency includes coding delay, air-interface latency, core network latency, processing delay, transmission delay, decoding delay, and display delay. For control data transmission, the end-to-end latency includes air-interface latency, core network latency, processing delay, and transmission delay. It is predicted that from the year 2020 onwards, the end-to-end latency for remote drone control should be less than 100 ms and the corresponding network latency (including air-interface and core network latency) should be less than 20 ms.

#### C. HD Image/Video Transmission

HD image/video transmission can greatly expand the application scenarios of drones including power line and BS inspection, agriculture exploration, command & rescue, entertainment, and surveillance. With high data rate transmission, drones connected to mobile networks will be able to transmit HD images/videos to support augmented reality (AR) and virtual reality (VR) immersive experience, which can make remote control more effective and accurate.

To enable HD image/video transmission, mobile networks should be able to provide high uplink (from drone to BS) data rate for drones. The required data rate is mainly determined by image/video size and its quality. For example, with 1080p digital image transmission quality (1920x1080), the required data rate is about 4 Mbps. In the future, with the increased demand of higher resolution images/videos in vertical industries that need the support of 4K/8K HD video and AR/VR services, higher data rate at the Gbps level will be needed. 5G networks are well positioned to support such services with multi-Gbps data rate requirement.

#### D. Drone Identification and Regulation

Mobile networks can facilitate drone identification and regulation by assisting with drone registration, monitoring, forecast and coordination.

- **Registration:** Through defining and standardizing drone equipment number, serial number, drone IMEI number, and flight control serial number, the whole process from initial drone production to in-use can be monitored orderly. Through the standardization of registration process of drone users and owners and mobile networks assisted verification, drone users and owners can be legally supervised.
- **Monitoring:** Drones connections and data communications via mobile networks can be identified and monitored. With additional regulatory protocols, drone applications can be fully monitored real time.
- **Forecast:** By tracking drone locations and monitoring flight traffic and route, flight conditions can be dynamically assessed, and early warning of potential risks can be made accordingly.
- **Coordination:** Through the authorized supervision of all the involved vertical industries, information sharing among industries and different enterprises can be

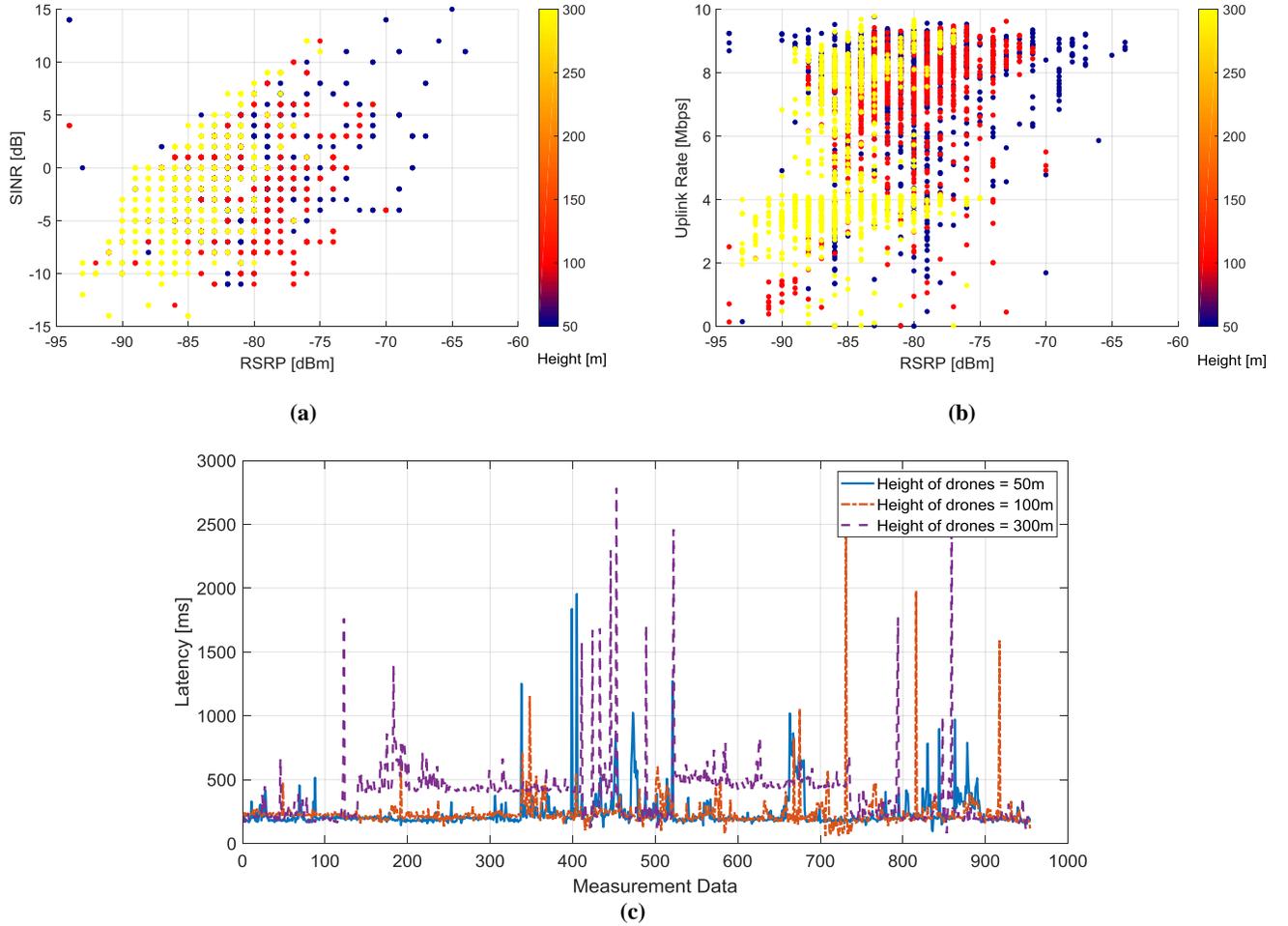

**Figure 2: Field measurement results in a CMCC 4G+ network: (a) scatterplot of SINR vs. RSRP; points are colored by height (see color bar scale), (b) scatterplot of uplink rate vs. RSRP; points are colored by height (see color bar scale), (c) latency performance at different heights**

coordinated, leading to more efficient and effective drone traffic control.

*E. High-precision Positioning*

Positioning is important for various drone applications. Besides conventional positioning in the horizontal plane, vertical positioning is also critical in many drone applications. With the development of drone applications, the requirement on positioning accuracy may increase from dozens of meters to sub-meter level.

Table 1 summarizes four levels of positioning accuracy requirements. For normal monitoring tasks, 50 m positioning accuracy is sufficient. Applications such as farmland mapping and auto-charging require sub-meter level high-precision positioning.

IV. LTE-ADVANCED CONNECTED DRONES FIELD TRIAL RESULTS

In this section, we present field trial results collected in a commercial LTE-Advanced network to gain insights into the capabilities of 4G+ networks for connected drones.

The field trials were carried out with a drone equipped with data acquisition module, a flight control device, and a data processing server. The measurement results were collected in a LTE-Advanced network at 2.6 GHz carrier frequency with 20 MHz carrier bandwidth. Data were collected at different heights and sent to the data processing center using the LTE-Advanced network.

In Figure 2(a), we present a scatterplot of SINR versus RSRP, where points are colored by height. In Figure 2(b), we present a scatterplot of uplink rate versus RSRP, where points are colored by height. In Figure 2(c), we present the latency performance at three different heights: 50m, 100m, and 300m.

The results in Figures 2(a) show a declining trend of downlink SINR and RSRP as the drone height increases. The RSRP values are mainly concentrated between -75 dBm and -90 dBm, which implies that the coverage for drones at 50-100 m is satisfactory. However, as the height increased, the pathloss becomes larger and the BS antenna gain reduces, which leads to reduced received power. The SINR values are mainly concentrated between -10 dB and 5 dB, which implies the interference for drones in the air is strong. This is because that the propagation becomes closer to LOS free-space as the height increases. As a result, stronger interfering signals from neighboring cells are received at the drone UE, which leads to reduced SINR. The results show that the coverage of the measured LTE-Advanced network for drones can be up to 100

|  | Altitude Coverage | | Wide area Coverage | |
|---|---|---|---|---|
|  | Height | Typical use case | Scenario | Typical use case |
| Coverage | 10 m | Vegetation protection (spraying of agricultural chemicals) | Hotspot area (Stadium, tourist area, commercial area, farmland) | Aerial entertainment, Agriculture inspection |
|  | 50-100 m | Powerline/BS inspection, Rescue, Aviation entertainment, Aerial monitoring, Logistics | Along-the-line area (Power station, BS tower) | Powerline/BS inspection |
|  | 200-300 m | Mapping of farmland information | Urban Macro area | Rescue, Aerial monitoring |
|  | 300-3000 m | Upper-air inspection (e.g. pipeline) | Urban, Suburban and Rural | Logistics and transportation |
|  | Level | Value | Typical application | |
|  | 1 | Uplink 200 kbps | Control and command transmission | |
| Data rate | 2 | Uplink 4 Mbps | 1080p data transmission | |
|  | 3 | Uplink 15 Mbps | 4K HD video | |
|  | 4 | Uplink 60 Mbps | 8K HD video | |
|  | 5 | Uplink 1 Gbps | AR/VR | |
|  | Level | End-to-end latency | Network latency | Typical application |
| Latency | 1 | < 400ms | < 40ms | Image/video transmission |
|  | 2 | < 100ms | < 20ms | Remote real-time control |
|  | Level | Accuracy | Typical application | |
|  | 1 | < 50m | Aerial surveillance | |
| Positioning | 2 | < 10m | Flight control at the current stage | |
|  | 3 | < 1m | Flight control at a future stage | |
|  | 4 | 0.1 m | Mapping of farmland, Automatic charging | |

**Table 1: Connectivity requirements of drone uses and applications**

| Frequency | Bandwidth | Peak Rate for Single User | Cell Radius | Edge Rate for Single User |
|---|---|---|---|---|
| 3.5 GHz | 100 MHz | DL:1.3 Gbps [4 TR, 256 QAM] UL: 175 Mbps [2 TR, 64 QAM] | 300 m | DL: 200 Mbps, UL: 4 Mbps |
|  |  |  | 200 m | DL: 450 Mbps, UL: 16 Mbps |
|  |  |  | 100 m | DL: 650 Mbps, UL: 40 Mbps |
| 26 GHz | 1 GHz | DL: 6.5 Gbps/13 Gbps [2 TR/4 TR, 256 QAM] UL: 1.75 Gbps [2 TR, 64 QAM] | 50 m | DL: 5 Gbps/10 Gbps [2 TR/4 TR] UL: 200 Mbps [2 TR] |

**Table 2: Expected 5G data rates according to different frequency bands and configurations**

m. This coverage level can meet the requirements of many drone applications. However, the measured LTE-Advanced network has difficulty in providing ubiquitous coverage for drones flying above 100 m.

From the scatterplot of uplink rate versus RSRP in Figure 2(b), it can be observed that the data rate performance can sufficiently satisfy the command and control transmission rate requirement of 200 kbps as well as the 1080p image transmission rate requirement of 4 Mbps most of the time. However, the measured LTE-Advanced network was not able to meet the transmission rate requirement of high definition images/videos (such as 8K high definition video), which would require higher than 10 Mbps date rate.

The latency in Figure 2(c) was measured based on ping package (32 bytes) from the drone to the LTE-Advanced network and back to the drone, which was the end-to-end latency including air-interface latency, core network latency, processing delay, and transmission delay. The results show that most of the latency data samples are concentrated between 200 ms and 300 ms at the height of 50 m or 100 m. When the drones fly up to 300 m, the latency data samples are concentrated between 400 ms and 500 ms. These results suggest that without further enhancements the measured LTE-Advanced network might have difficulty in serving drone applications that require stringent latency performance (such as 100 ms).

The presented field trial results illustrate that the coverage, data rate, and latency in the existing LTE-Advanced network may need to be improved and upgraded to better serve the connected drones. With more advanced technologies, 5G mobile networks have more powerful capabilities to fulfill these objectives, as discussed in the next section.

## V. CAPABILITIES OF 5G FOR DRONES

The improved capabilities of 5G networks have the potential to provide efficient and effective mobile connectivity for large-scale drone deployments with more diverse uses. In particular, 5G networks will introduce new technologies to provide 3D coverage enhancement, high data rate, customized end-to-end QoS guarantee, and efficient identification and monitoring based on big data analysis.

### A. Coverage Enhancements and High Date Rate

The current mobile communication networks belong to the Public Land Mobile Networks (PLMN), which is mainly designed for terrestrial terminals. The flight height of a drone is usually close to or higher than the height of BS antenna. Due to the LOS propagation and sidelobe coverage from neighboring cells, the radio environment of drones is significantly different from that of terrestrial terminals [5][7]. Drones are more likely to observe more interferences [13], cell coverage irregularities, and complex neighbor cell relationship than conventional terrestrial users, which lead to mobility management challenges for drones and thus more complexity for aerial coverage.

5G networks can provide solutions to the above problems through the introduction of new technologies such as massive multiple-input and multiple-output (MIMO) techniques, enhancements of coordinated Multi-Point transmission (CoMP), automatic neighbor cell relationship, optimization of measurement reports, and optimized power control mechanisms [14]. For example, dedicated or optimized cell sites or antennas can be used for aerial coverage; 3D beamforming can be used to achieve directional signal transmission or reception; and different frequencies can be used to isolate the interference of different types of users. Neighbor cell coverage and mobility management strategies can be customized to achieve coverage optimization in the skies. Information such as flight route and flight status of the drones can be exploited for connectivity management and optimization.

With wide bandwidths and massive MIMO, 5G can provide various levels of downlink and uplink high data rates at different frequency bands with different configurations, as illustrated in Table 2.

### B. Service Differentiation and QoS Enhancements

In a 5G network, an end-to-end QoS system consisting of new technologies such as network slicing and two-level QoS mapping has the potential for better meeting the needs of increasingly diverse drone applications. Network slicing allows multiple logical networks to be created on top of a common shared physical infrastructure [15]. Each slice is a complete logical network consisting of network capabilities and the associated resources which provide specific end-to-end service capabilities. The slices of the network are isolated from each other in the control and user planes. By allocating a dedicated slice for the connected drones, it is possible to logically distinguish the service and radio resource management for the drones from those for the terrestrial terminals. By incorporating drone application specific information, it is possible to provide service differentiation for different drone uses by for example using different slices to support control signaling and different application data services. With network slicing, the physical network is partitioned at an end-to-end level to allow optimized grouping of drone traffic such as low latency drone traffic and high data rate drone traffic, and to isolate drone traffic from terrestrial traffic of different characteristics.

### C. Detection and Resource Management

Mobile networks should recognize a drone UE to apply drone UE specific radio resource management methods and achieve the right service optimization for the drone UE. One option is to assign special subscriber profile identity to identify the drone UE. Another option could be to introduce direct indication mechanisms so that drone UEs will inform the network when they are in the flying mode.

Apart from relying on explicit identifiers or direct indication, it is also possible to identify a drone-mounted UE through proprietary algorithms. These proprietary solutions are of special interest for identifying a "rogue" drone that carries a terrestrial UE. These rogue drones are being observed in the field where it is found that some users may attach their mobile devices that do not have "drone capability" to drones and then fly the drones [8]. These rogue drones may generate excessive interference to the network and may not be allowed by regulations in some regions. Identifying a rogue drone may enable the network to take proper measures. For example, drone UE specific radio resource management methods may be used if the network detects the rogue drone UE but decides to continue serving the UE. Alternatively, the network may limit the service or even drop the connection. Mobile network operators may provide rogue drone detection services to assist with drone traffic management, law enforcement, or other business purposes such as insurance.

To ensure effective implementation of the drone-specific network optimization solutions and to meet drone regulations such as flight restrictions, 5G networks will be better positioned to effectively identify and monitor drones. Self-organized network (SON) realizes network intelligence with self-organizing capabilities. Through centrally collecting and mining big data information of key performance indicators and network configuration parameters, SON may implement system-level, BS-level, and terminal-level information collection and decision-making. With the capabilities of big data collection and analysis, 5G SON will enable more efficient drone identification and management. Specifically, with the collection and analysis of historical network performance data, airborne radio environment distribution models can be constructed, and drone UE can be accurately identified by machine learning or pattern recognition methods. To assist with drone regulations and air traffic management, the big data collected by a 5G network can be used for flight path planning to avoid coverage holes, congestion, collisions, and no-fly zones.

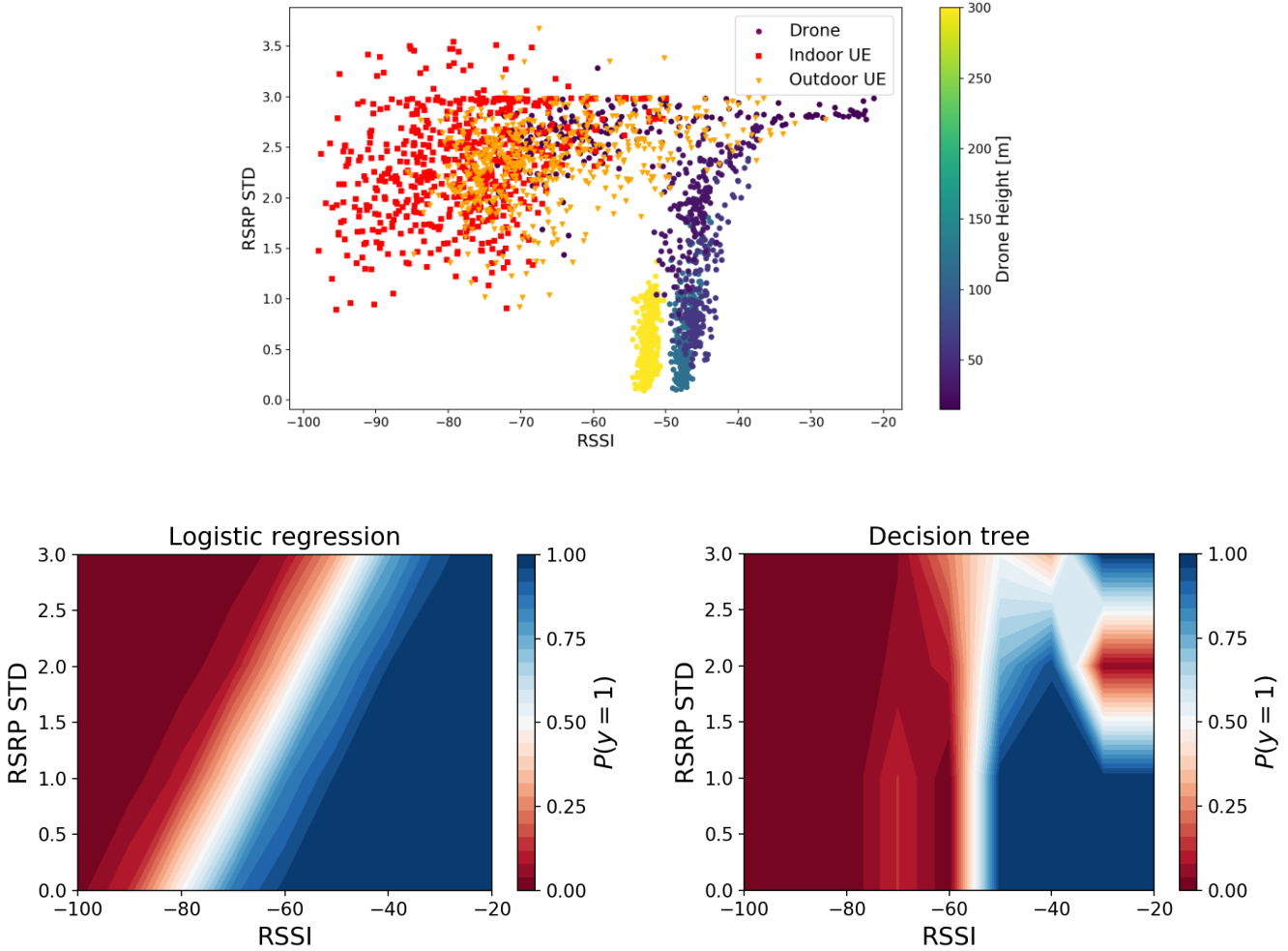

**Figure 3: An example of drone identification using radio measurements in mobile networks**

Before ending this section, we give a concrete example to illustrate how some of the collected data can be exploited by 5G networks using machine learning methods to assist with drone identification. The example is taken from a simulated environment with BSs deployed with 500 m inter-site distance. The example is illustrated in Figure 3. In this example, we utilize two types of radio measurement data: received signal strength indicator (RSSI) and standard deviation of the eight strongest RSRPs (RSRP STD). In the top subfigure in Figure 3, each square represents a data sample from an indoor UE, each triangle for an outdoor UE, and each dot for a drone UE. The data samples for drone UEs at different heights are marked according to the color bar on the right side of the top subfigure in Figure 3. It can be seen that the separation of the data samples for drone UEs from the data samples for terrestrial UEs increases with the altitude. We then apply two classification machine learning models, Logistic Regression and Decision Tree, using the two features: RSSI and RSRP STD to identify if a UE is a drone UE or not. The two bottom subfigures in Figure 3 show the probability regions with the trained Logistic Regression and Decision Tree models in the investigated scenario, respectively. The figure shows how the Decision Tree model has classified the region with high RSSI and medium-to-high RSRP STD (the upper right region in the bottom right subfigure of Figure 3) as low drone UE probability region, while the Logistic Regression model has classified the corresponding region as high drone UE probability region.

## VI. CONCLUSIONS AND RESEARCH DIRECTIONS

The drone ecosystem can greatly benefit from mobile networks that offer a variety of capabilities and features for identifying, tracking, and controlling the growing fleet of drones. Mobile networks provide wide-area, quality, and secure connectivity for drone operations beyond visual LOS. In this article, we have analyzed drone communications requirements on mobile networks in terms of coverage, rate, latency, and positioning. We present field trial results collected in LTE-Advanced networks to shed light on the capabilities of the current 4G+ networks for supporting the internet of drones. We find that the existing LTE-Advanced networks targeting terrestrial usage can already support the initial deployment of low altitude drones. The significantly improved capabilities of 5G networks will provide more efficient and effective mobile connectivity for large-scale drone deployments with more diverse applications. We firmly believe that mobile technologies, based on evolving global standards, will be the essential foundation for the vibrant global growth of the drone ecosystem and pave the way for making the internet of drones a reality.

Internet of drones is an emerging and underexplored field. We conclude by pointing out some fruitful avenues pertinent to mobile networks for future research.

**Connectivity from the sky:** We have focused in this article on providing connectivity to the sky, where drones are UEs. If the communication infrastructure is damaged due to natural disasters or not available in under-developed or rural areas, drones can provide mobile connectivity by becoming BSs or relays. One prominent example is the Facebook's large solar-powered high-altitude Aquila drone that is aiming to connect areas that are off the grid to the internet. How to integrate the supplemental airborne connectivity with the existing mobile networks deserves further study.

**Interaction with drone traffic management systems:** Conventional air traffic management systems are mainly designed for manned aircrafts and will likely not be scalable to cover the expected high density of low altitude drones. New unmanned aircraft systems traffic management (UTM) systems are being developed to safely handle the likely high density of low altitude drone traffic. One prominent example is the UTM architecture proposed by National Aeronautics and Space Administration (NASA) and the Federal Aviation Administration (FAA) in the US. Mobile networks can add valuable services to the UTM systems such as independent drone identification, authorization, and backbone communications for information distribution. The integration of mobile networks with the UTM systems is a largely underexplored field of immediate practical relevance.